\documentclass[paper]{revtex4}

\usepackage{graphicx}
\usepackage{axodraw}
\usepackage{epsfig}
\usepackage{fancyhdr}
\usepackage{empheq}
\usepackage{mathbbol}
\usepackage{wasysym}
\usepackage{pstricks}
\usepackage{color}
\usepackage{bbm}
\def\qq{\mathbbmss{q}}

\def\qu{{q_{_1}}}
\def\qd{{q_{_2}}}
\def\qub{{\bar q_{_1}}}
\def\qdb{{\bar q_{_2}}}

\begin{document}

\title{Exotic Hadrons with Hidden Charm and Strangeness}

%

\author{NV Drenska$^{\dag,*}$, R Faccini$^{\dag, *}$} 
\author{AD Polosa$^*$}
\affiliation{$^\dag$Dipartimento di Fisica, Universit\`a di Roma `La Sapienza', Piazzale A Moro 2, Roma, I-00185, Italy\\
$^*$INFN Roma, Piazzale A Moro 2, Roma, I-00185, Italy}

\begin{abstract}
We investigate on exotic tetraquark hadrons of the kind $[cs][\bar{c}\bar{s}]$ by 
computing their spectrum and decay modes within a constituent diquark-antidiquark model. 
We also compare these
predictions with the present experimental knowledge.
 
\end{abstract}

\maketitle

\thispagestyle{fancy}

{\bf \emph{Introduction}}. In the last few years we are witnessing the discovery of a number of new narrow hadronic resonances with charm
which do not match standard quark-antiquark interpretation, thereby named exotic hadrons. This has called for alternative interpretations of 
their inner structure. One of the possible explanations is that these particles are loosely bound molecules of open charm mesons~\cite{mol}. 
Another possibility is that new aggregation patterns of quarks in matter are possible. We follow the suggestion by Jaffe and Wilczek of having di-quarks as 
building blocks~\cite{jw}. Light diquarks have been object of several lattice studies. Recently we have studied the hypothesis of treating 
the diquark as a fundamental field in large $N$ Chromodynamics in two dimensions~\cite{gjp} finding some qualitative features of the spectra which are 
common to real ones. The idea that the colored diquark can be handled as a constituent building block is at the core of the approach taken in this paper. 

Motivated by our former study on the interpretation of the $Y(4260)$ resonance~\cite{y4260}, we analyze the possibility of a spectroscopy of particles with hidden strangeness and charm
embodied in  diquark-antidiquark structures of the kind $[cs][\bar c\bar s]$, where $\qq=[cs]$ is a ${\bf \bar 3_c}$ diquark. We predict the mass spectrum for these states and discuss which might be their
prominent decay modes on the basis of quark rearrangements in the $\qq\bar \qq$ system.
The mass spectrum is computed using the non-relativistic spin-spin interactions 
Hamiltonian, supposed to remove the degeneracy among the various $[cs][\bar c\bar s]$ states
with assigned spins and anguar momenta
\begin{equation}
H_{_{SS}}=\sum_{\rm pairs} \frac{\kappa_{ij}}{m_i m_j}\,(\vec S_i \cdot \vec S_j)\, \delta^{3}(\vec r_{ij})
\label{eq:dgg}
\end{equation}
The couplings  are inversely proportional to quark masses: we will incorporate this dependency in the color-magnetic moments $\kappa$. 
The Hamiltonian~(\ref{eq:dgg}) describes  contact interactions. For this reason 
we expect that allowing a relative orbital angular momentum  between the diquarks will decrease or switch-off the  spin-spin interactions between quarks and antiquarks. We will take into account this effect
in  the determination of the spectrum of the negative parity particles. 

{\bf \emph{The model}}. 
Adopting the approach discussed at length  in~\cite{xmppr} and~\cite{drenska} 
we will determine the mass spectrum of  $[cs][\bar{c}\bar{s}]$ hadrons by diagonalization of the following non-relativistic effective Hamiltonian for a $[q_1q_2][\bar q_1\bar q_2]$ diquark-antidiquark hadron
\begin{equation}
H=2m_\qq + H_{_{SS}}^{(qq)} + H_{_{SS}}^{(q\bar q)}+H_{_{SL}}+H_{_{L}} 
\label{ham0}
\end{equation}
where
\begin{eqnarray}
&&H_{_{SS}}^{(qq)}= 2\kappa_\qq 
(\vec S_\qu\cdot \vec S_\qd+ \vec S_{\qub}\cdot \vec S_{\qdb}) \notag\\ 
&&H_{_{SS}}^{(q\bar q)}=  2\kappa_{\qu\qdb}(\vec S_\qu\cdot \vec S_{\qdb}+
\vec S_{\qub}\cdot \vec S_\qd)+2\kappa_{\qu\qub} \vec S_\qu\cdot \vec S_{\qub}
+2\kappa_{\qd\qdb} \vec S_\qd\cdot \vec S_{\qdb}
\label{hams}
\end{eqnarray}
represent the chromomagnetic interactions between quarks in the tetraquark system whereas the spin-orbit and orbital contributions are given by
\begin{eqnarray}
&&H_{_{SL}}= 2A (\vec S_\qq\cdot\vec L+\vec S_{\bar \qq}\cdot\vec L)\notag\\
&&H_{_{L}}= B \frac{L(L+1)}{2}
\label{haml}
\end{eqnarray}
respectively.
The symbol $\vec S_\qq$ represents the {\it total} spin of the diquark $\qq = [q_1 q_2]$. $A,B$ are coefficients to be determined by data.
To diagonalize $H$ we need to estimate the diquark mass $m_\qq$ and the coupling constants $\kappa$; then we have to  specify the states having assigned $J^{PC}$ quantum numbers and find their masses.
The $[cs]$ diquark mass is simply estimated by  $m_{[cs]}=m_{[qs]}+m_c - m_q=1955~\rm{MeV}$,
where $m_{c,q}$ are constituent masses and $m_{[qs]}$ is obtained by comparison with light scalar mesons data~\cite{thooft}.  
We have used $m_{[qs]}=590$~MeV, $m_c=1670$~MeV, $m_q=305$~MeV.

As shown in~\cite{xmppr}, quark-antiquark  spin-spin couplings are 
estimated by comparison with the mass spectra of ordinary $q\bar{q}$ mesons. 
A calculation made along these lines provides us with the  
chromomagnetic couplings for $q\bar q$ color singlets $(\kappa_{q\bar q})_{\bf 1}$.
We obtain $\kappa_{cs}=25$~MeV, $(\kappa_{c\bar s})_{\bf 1}=72$~MeV, $(\kappa_{s\bar s})_{\bf 1}=121$~MeV, $(\kappa_{c\bar c})_{\bf 1}=59$~MeV,
$A=22$~MeV, $B=495$~MeV.

On the other hand, the couplings $\kappa_{q\bar q}$ in~Eq.~(\ref{hams})
are not necessarily in the singlet channel as those estimated 
since octet couplings ($\kappa_{\bf 8}$) are also possible between quarks and antiquarks 
in a $\qq\bar\qq$ system~\footnote{A quark $q$ in the diquark $\qq$ could have a color octet spin-spin interaction with an antiquark $\bar q^\prime$ in the antidiquark $\bar{\qq}^\prime$. Same for the remaining quark-antiquark pair to get a singlet by ${\bf 8}\otimes {\bf 8}$.}. The octet couplings are estimated with the aid of the one-gluon exchange model as follows. For the diquark attraction is in the ${\bf \bar 3}$-color channel, we can write $\qq^i=[cs]^i:=\epsilon^{ijk}c_j s_k$, neglecting spin. $i,j,k$ are color indices in the fundamental representation of $SU(3)$. Then the color neutral hadron is
\begin{equation}
[cs][\bar{c}\bar{s}]
=\epsilon^{ijk}\epsilon_{ij^\prime k^\prime}(c_{j}s_{k})({\bar{c}}^{j^\prime}{\bar{s}}^{s^\prime})
= (c_{j}{\bar{c}}^{j})(s_{k}{\bar{s}}^{k}) - (c_{j}{\bar{s}}^{j})(s_{k}{\bar{c}}^{k})
\label{eq:color}
\end{equation}
We then use the following  $SU(N)$ identity for the Lie algebra generators
\begin{equation}
\sum_{a=1}^{N^2-1}\lambda_{ij}^{a}\lambda_{kl}^{a}
=2\left (\delta_{il}\delta_{jk}-\frac{1}{N}\delta_{ij}\delta_{kl} \right ) 
\end{equation}
where $N$ is the number of colors.
A color octet (N=3) $q\bar q$ state can be written as $\bar q^i \lambda^a_{ij} q^j$, and consequently
\begin{equation}
({\bar{c}}^{i}\lambda_{ij}^{a}{c}^{j})({\bar{s}}^{k}\lambda_{kl}^{a}{s}^{l}) 
= \sum_{a}\lambda_{ij}^{a}\lambda_{kl}^{a} \bar{c}^{i}c^j \bar{s}^k s^l
=2\left [(c_i{\bar{s}}^i)(s_k{\bar{c}}^k) -\frac{1}{N}(c_i{\bar{c}}^i)(s_k{\bar{s}}^k)\right ]
\label{eq:clcsls}
\end{equation}
This allows to extract from~(\ref{eq:color}) the octet term as follows:
\begin{equation}
[cs][\bar{c}\bar{s}]
=(c_{j}{\bar{c}}^{j})(s_{k}{\bar{s}}^{k}) -
\left [\frac{1}{2} ({\bar{c}}^{i}\lambda_{ij}^{a}{c}^{j})({\bar{s}}^{k}\lambda_{kl}^{a}{s}^{l})
+\frac{1}{3}({c}_{j}{\bar{c}}^{j})(s_{k}{\bar{s}}^{k})\right ]
= \frac{2}{3}({c}_{j}{\bar{c}}^{j})(s_{k}{\bar{s}}^{k})
-\frac{1}{2}({\bar{c}}^{i}\lambda_{ij}^{a}{c}^{j})({\bar{s}}^{k}\lambda_{kl}^{a}{s}^{l})
\end{equation}
This formula gives information about the relative weights of a singlet and an octet color state in a diquark-antidiquark tetraquark. 
We have three colors running in the sum $c_i {\bar{c}}^i$ whereas $a=1,...,8$ in 
${\bar{c}}^{i}\lambda_{ij}^{a}{c}^{j}$. Therefore the probability of finding (projecting onto)  a particular 
$q\bar q$ pair in color singlet, for example to find $c\bar c$ in the color singlet state $c_j\bar c^j$,
is half the probability of finding the same pair in color octet $\bar c \lambda^a c$ as $3\times 2/3 =1/2(8\times 1/2)$. We write then:
\begin{equation}
\kappa_{c\bar{c}} ([cs][\bar{c}\bar{s}])=\frac{1}{3}(\kappa_{c\bar{c}} )_{\bf{1}} 
+ \frac{2}{3}(\kappa_{c\bar{c}} )_{\bf{8}} 
\label{eq:superposition}
\end{equation}
where $(\kappa_{c\bar{c}} )_{\bf{1}} $ have been reported above.
For the determination of the quantity $(\kappa_{c\bar{c}} )_{\bf{8}}$ we have to resort to the one-gluon exchange model.
In this model we assume that the coupling $\kappa_{q q(\bar q)}({\bf R})$, ${\bf R}$ being the color representation of the two quark system  can be considered  proportional to the product of the two color charges in the vertices of a $q q(\bar q)\to qq(\bar q)$ diagram in which one gluon is exchanged between quarks. No distance dependency is taken into account.

Then, if $\kappa_{c\bar{c}}({\bf{R}})$ is the weight of the quark-antiquark interaction, writing the above mentioned product of color charges in terms of  the $SU(3)$ Casimir operators we have  
\begin{equation}
\kappa_{c\bar{c}}({\bf{R}}) \sim \left ( C^{(2)}({\bf{R}}) - C^{(2)}({\bf{3}}) -C^{(2)}({\bf{\bar{3}}}) \right )
\label{eq:casimir}
\end{equation}
We recall that  $C^{(2)}({\bf{R}})=0,4/3,4/3,3$ as ${\bf{R}}={\bf{1}},{\bf{3}},{\bf{\bar{3}}},{\bf{8}}$. Then it is immediately found that 
\begin{equation}
(\kappa_{c\bar{c}})_{\bf{1}}  \sim   - \frac{8}{3} \qquad 
(\kappa_{c\bar{c}})_{\bf{8}}   \sim  \frac{1}{3}  = -\frac{1}{8} (\kappa_{c\bar{c}})_{\bf{1}} 
\label{eq:cas18}
\end{equation}
Finally, from Eq.~(\ref{eq:superposition}), we have
\begin{equation}
\kappa_{c\bar{c}}([cs][\bar{c}\bar{s}]) = \frac{1}{4} (\kappa_{c\bar{c}})_{\bf{1}}
\label{eq:onefour}
\end{equation}

Now that we know its input parameters, we are ready to diagonalize the Hamiltonian~(\ref{ham0}).
We label the particle states by using the notation $|S_\qq, S_{\bar \qq};S_{ \qq\bar\qq}, J \rangle $ 
where $S_{ \qq\bar\qq}$ is the total spin of the diquark-antidiquark system.
States are organized in order to have definite $J^{PC}$ quantum numbers. 
For negative parity ones a unit of relative angular momentum between the diquark and 
the antidiquark is required  ($L_{\rm \qq\bar\qq} =1$).
Altogether we have

{\bf \emph{a-}}Two positive parity states states with $J^{PC}$ = $0^{++}$
\begin{equation}
{|0^{++}\rangle}_1 =|0_{cs}, 0_{\bar{c}\bar{s}};0_{\rm \qq\bar\qq}, J=0 \rangle \qquad
{|0^{++}\rangle}_2 =|1_{cs}, 1_{\bar{c}\bar{s}};0_{\rm \qq\bar\qq}, J=0 \rangle 
\end{equation}

{\bf \emph{b-}}Three states with $J=0$ and negative parity ($L_{\rm \qq\bar\qq} =1$ required)
\begin{equation}
|A\rangle =|1_{cs}, 0_{\bar{c}\bar{s}};1_{\rm \qq\bar\qq}, J=0 \rangle  \qquad
|B\rangle =|0_{cs}, 1_{\bar{c}\bar{s}};1_{\rm \qq\bar\qq}, J=0 \rangle  \qquad
|C\rangle =|1_{cs}, 1_{\bar{c}\bar{s}};1_{\rm \qq\bar\qq}, J=0 \rangle 
\end{equation}
State $|C\rangle$ is even under charge conjugation. 
Taking  symmetric and antisymmetric combinations of states $|A\rangle $ 
and $|B\rangle $ we obtain a {\cal{C}}-odd and a {\cal{C}}-even state respectively; therefore we have
two states with $J^{PC}=0^{-+}$ 
\begin{equation}
{|0^{-+}\rangle}_1 =\frac{1}{\sqrt{2}} \left (|A \rangle - |B \rangle \right ) \qquad
{|0^{-+}\rangle}_2 =|C\rangle 
\end{equation}
and one state with $J^{PC}=0^{--}$
\begin{equation}
|0^{--}\rangle = \frac{1}{\sqrt{2}} \left ( |A \rangle + | B \rangle \right )
\end{equation}

{\bf \emph{c-}}Three states with $J=1$ and positive parity
\begin{equation}
|D\rangle =|1_{cs}, 0_{\bar{c}\bar{s}};1_{\rm \qq\bar\qq}, J=1 \rangle  \qquad
|E\rangle =|0_{cs}, 1_{\bar{c}\bar{s}};1_{\rm \qq\bar\qq}, J=1 \rangle  \qquad
|F\rangle =|1_{cs}, 1_{\bar{c}\bar{s}};1_{\rm \qq\bar\qq}, J=1 \rangle 
\end{equation}
$|F\rangle$ is an eigenvector under charge conjugation, with negative eigenvalue. 
Operating on  $|D\rangle$ and $|E\rangle$ in the same way as 
for states $|A\rangle$ and $|B\rangle$ we obtain the $J^{PC}=1^{++}$ state
\begin{equation}
|1^{++}\rangle = \frac{1}{\sqrt{2}} \left ( |D \rangle + | E \rangle \right )
\end{equation}
and the $J^{PC}=1^{+-}$ ones
\begin{equation}
{|1^{+-}\rangle}_1 =\frac{1}{\sqrt{2}} \left (|D \rangle - |E \rangle \right ) \qquad
{|1^{+-}\rangle}_2 =|F\rangle 
\end{equation}

{\bf \emph{d-}}Six states with $J=1$ and negative parity. 
To start with consider the following two
\begin{equation}
|G\rangle =|1_{cs}, 0_{\bar{c}\bar{s}};1_{\rm \qq\bar\qq}, J=1 \rangle \qquad
|H\rangle =|0_{cs}, 1_{\bar{c}\bar{s}};1_{\rm \qq\bar\qq}, J=1 \rangle  
\end{equation}
differing from $|D\rangle $ and $|E\rangle $ as we have $L_{\rm \qq\bar\qq}=1$ here. 
When symmetrized and antisymmetrized these give the combinations
\begin{equation}
{|1^{-+}\rangle}_1 =\frac{1}{\sqrt{2}} \left (|G \rangle - |H \rangle \right )\qquad
{|1^{--}\rangle}_1 =\frac{1}{\sqrt{2}} \left (|G \rangle + |H \rangle \right )
\end{equation}
Moreover we have the following four  charge conjugation eigenstates
\begin{eqnarray}
&&{|1^{-+}\rangle}_2 =|1_{cs}, 1_{\bar{c}\bar{s}};1_{\rm \qq\bar\qq}, J=1 \rangle  \qquad {|1^{--}\rangle}_2 =|0_{cs}, 0_{\bar{c}\bar{s}};0_{\rm \qq\bar\qq}, J=1 \rangle  \nonumber\\
&&{|1^{--}\rangle}_3 =|1_{cs}, 1_{\bar{c}\bar{s}};0_{\rm \qq\bar\qq}, J=1 \rangle  \qquad {|1^{--}\rangle}_4 =|1_{cs}, 1_{\bar{c}\bar{s}};2_{\rm \qq\bar\qq}, J=1 \rangle
\end{eqnarray}


The action of the  spin operators in Eq.~(\ref{hams}) on the states here listed  is independent of the specific $L_{\rm \qq\bar\qq}$ value. Let us write
\begin{equation}
|S_\qq,S_{\bar \qq};S_{\rm \qq\bar\qq}, J\rangle=|c^T\Gamma s, \bar c^T\Gamma \bar s; S_{\rm \qq\bar\qq}, J \rangle
\label{statess}
\end{equation}
where the $\Gamma$ can be
$\Gamma^0 =1/ \sqrt{2}\; \sigma_{2}$ and $\Gamma^i = 1/\sqrt{2}\; \sigma_{2} \sigma^{i}$
for spin 0 and spin 1, respectively. The numerical factors are chosen in such a way to preserve  the normalization
$\rm{Tr}[{(\Gamma^\alpha )}^{\dag} \Gamma^\beta ] = \delta^{\alpha \beta} $.
Then the action of a spin-spin interaction operator, e.g. 
$\vec S_{ c}\cdot \vec S_s$, on the generic state in Eq.~(\ref{statess}) is:
\begin{equation}
(\vec S_{c}\cdot \vec S_s) |c^T\Gamma s, \bar c^T\Gamma \bar s; S_{\qq\bar\qq}, J \rangle =
\frac{1}{4}\sum_j |c^T\sigma_j^T \Gamma \sigma_j s, \bar c^T \Gamma \bar s; S_{\qq\bar\qq}, J \rangle
\label{eq:action}
\end{equation}
and similarly  for the other operators.

\begin{figure}[htb]
\begin{center}
\epsfig{height=5truecm, width=9truecm,figure=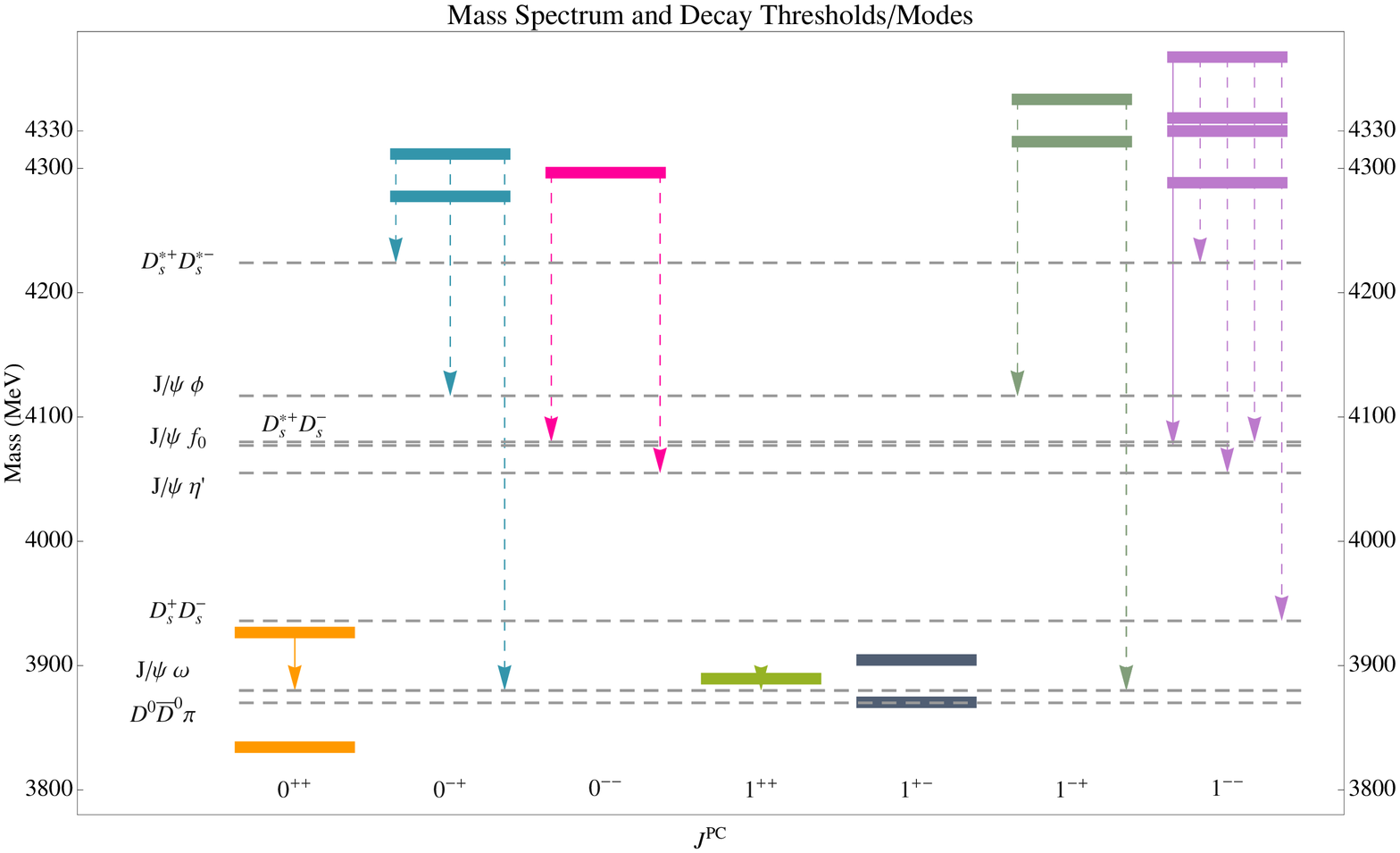}      
\caption{[$cs$][$\bar{c}\bar{s}$] spectrum and decay thresholds of prominent channels. Dashed lines are for $P-$wave decays, solid lines for
$S-$wave decays.  As commented in~\cite{y4260}, with the parameters at hand, we predict a mass value for the $Y(4260)$ of $m_Y=4330\pm 70$~MeV.}
\label{fig:thresh}
\end{center}
\end{figure}

{\bf \emph{Results and discussion}}. The final results on the mass spectrum determination are summarized in Tab.~\ref{tab:qnum} together with the  spin, orbital quantum numbers, decay modes and widths, when calculable. We include in parentheses the mass shifts in MeV
due to turning off the spin-spin interactions whence orbital angular momentum effectively increases 
the diquark-antidiquark distance.  
For a pictorial summary see also Fig.~\ref{fig:thresh}.

\begin{table}[htb]
\begin{center} 
\begin{tabular}{|c|c|c|c|c|c|c|c|}
\hline
$S_\qq$	& 	$S_{\bar \qq}$	&	$S_{\rm \qq\bar\qq}$	
&	$L_{\rm \qq\bar\qq}$		 &	$J^{PC}$ &	M(MeV) &  Decay Channel [$\Gamma_{\rm part}$(KeV)]& Relative Wave\\
\hline
\hline
0		&	0 		&	0		 &	0			 &	$0^{++}$ &	3834&  	-		&	\\
\hline
1		&	1 		&	0		 &	0			 &	$0^{++}$&	3927 &  ${\rm Multihadron}$ &	$-$\\
\hline
\hline
1		&	0 		&	1		 &	1			 &	$0^{-+}$ &	4277(+15)& $ J/\psi ~\phi$ [35],  $D_s^{*+} D_s^{*-}$ [10] &	$P$	\\
\hline
1		&	1 		&	1		 &	1			 &	$0^{-+}$ &	4312(+30) &  $J/\psi$ $\phi$ [46], $D_s^{*+} D_s^{*-}$ [24]&	$P$	\\
\hline
\hline
1		&	0		&	1 		 &	1 			 &	$0^{--}$  &	4297(-5)&$\psi$ $\eta(\eta')$ [245(110)], $D_s^+ D_s^{*-}$ [500]&	$P$\\ 
\hline
\hline
1		&	0 		&	1		 &	0			 &	$1^{++}$ &	3890  &	 ${\rm Multihadron}$ &	$-$\\ 
 \hline
 \hline
 1		&	0 		&	1		 &	0			 &	$1^{+-}$ &	3870&	$J/\psi$ $\eta$  [610] & $S$	\\
\hline
1		&	1 		&	1		 &	0			 &	$1^{+-}$ &	3905  & $J/\psi~\eta$  [650] &		$S$		\\
\hline
\hline
1		&	0 		&	1		 &	1			 &	$1^{-+}$ &	4321(+15) & $J/\psi ~\phi $ [52] &	$P$	\\
\hline
1		&	1 		&	1		 &	1			 &	$1^{-+}$ &	4356 (+30)&  $J/\psi$ $\phi$ [64]&	$P$\\
\hline
\hline
0		&	0 		&	0		 &	1			 &	$1^{--}$ &	4330 & $\psi ~\eta(\eta')$ [90(45)], $D_{s}^{(*)+}D_{s}^{(*)-}$ [27]; $J/\psi$ $f_{0}(980)$ &	$P$; $S$\\
\hline
1		&	0 		&	1		 &	1			 &	$1^{--}$ &	4341(-5) & $\psi$ $\eta(\eta')$[92(48)], $D_{s}^{(*)+}~D_{s}^{(*)-}$ [31]; $J/\psi$ $f_{0}(980)$ &	$P$; $S$\\
\hline   
1		&	1		&	0		 &	1			 &	$1^{--}$ &	4390(+40) & $\psi ~\eta(\eta')$ [100(58)], $D_{s}^{(*)+}D_{s}^{(*)-}$ [51]; $J/\psi$ $f_{0}(980)$&	$P$; $S$\\
\hline
1		&	1		&	2		 &	1			 &	$1^{--}$ &	4289(-41) & $\psi ~\eta(\eta')$ [83(36)],  $D_{s}^{(*)+}D_{s}^{(*)-}$ [13]; $J/\psi$ $f_{0}(980)$&	$P$; $S$\\
\hline
\hline
\end{tabular}\\[2pt]
\caption{Quantum numbers and masses for [$cs$][$\bar{c}\bar{s}$] states. We include in parentheses the mass shifts in MeV
due to turning off the spin-spin interactions whence orbital angular momentum effectively increases 
the diquark-antidiquark distance.  The dominant two-body decay channels are also indicated together with the relative angular momentum of the two produced particles and the partial width ($\Gamma_{\rm part}$) of the exchange diagram dominated decays. As for the estimated masses, we put in parentheses the corrections in MeV due to turning off the spin-spin interactions whence orbital angular momentum effectively increases the diquark-antidiquark distance. By the notation $D^{(*)}$ we mean $D$ or $D^{*}$.  
}
\label{tab:qnum}
\end{center}
\end{table}

Suppose that the tetraquark system could be described energetically by a double-well potential with the two light quarks lying in the two wells induced by the charm quarks, which can be considered as static  color  sources. The potential barrier separating the two wells prevents a quark in the diquark to bind with an antiquark in the antidiquark (and vice-versa). This process occurs anyways  at the rate of  the barrier penetration. We shall assume that this is the case for the quark passing anyhow through the barrier to bind with the antiquark in the other well, as represented in the following diagram
\vspace{1truecm}
\begin{center}
\SetScale{0.5}
\SetOffset(195,5)
\fcolorbox{white}{white}{
  \begin{picture}(318,109) (193,-133)
    \SetWidth{0.8}
     \Text(90,-13)[l]{$c$}
     \Text(90,-27)[l]{$ s$}
      \Text(90,-51)[l]{$\bar c$}
       \Text(90,-65)[l]{$\bar s$}
        \Text(69,-19)[l]{$\qq$}
        \Text(69,-57)[l]{$\bar\qq$}
    \Text(260,-40)[l]{$\psi X_{ss}$}  
    \ArrowLine(193,-27)(507,-27)
    \ArrowLine(195,-55)(318,-54)
    \Line(318,-54)(390,-102)
    \ArrowLine(390,-102)(509,-102)
    \ArrowLine(510,-56)(391,-56)
    \ArrowLine(316,-101)(198,-101)
    \ArrowLine(511,-131)(195,-131)
    \Line(391,-56)(316,-101)
  \end{picture}
}
\end{center}
\vspace{-2truecm}
where $X_{ss}=\phi,\omega,\eta, \eta^\prime$.  The charm quarks have no other choice than neutralize color in a charmonium meson. An alternative process is the formation of open charm mesons by rearranging of the strange quarks. 
We shall suppose that the latter two alternatives occur at almost the same rate (both of them pay the same energetic price of breaking the diquark bindings).

As for the decay widths, consider for example the  decay into $J/\psi \phi$ of the $0^{-+}$ particle with mass $M=4277$~MeV in Table~\ref{tab:qnum}. The $S$-matrix element is
\begin{equation}
\langle J/\psi(\eta,p^\prime)\phi(\epsilon,q)|Y_{4277}(p)\rangle={\cal G} \epsilon^{\mu\nu\rho\sigma} \eta_\mu\epsilon_\nu p_\rho q_\sigma 
\end{equation} 
where ${\cal G}$ must have dimensions of the inverse of a mass to let  the width have dimensions of energy. The decay at hand is a $P-$wave decay, therefore
\begin{equation}
\Gamma(Y_{4277}\to J/\psi \phi)=\frac{1}{3}\frac{A^2}{M_Y^2}\left(\frac{1}{8\pi M_Y^2}p^{*3}\right)
\end{equation}
where $p^*$ is the decay momentum in the reaction and we assumed ${\cal G}=g/M_Y$, the dimensionless $g$ being written as $g=A/(\sqrt{2}M_Y)$. 
Assuming that the exchange diagrams in $[c(u,d)][\bar c \bar (\bar u,\bar d)]$ states have the same amplitudes as the ones in $[cs][\bar c\bar s]$~\cite{xmppr}, {\it i.e.}, assuming 
$A=2.6$~GeV, we obtain that the decay modes reported in Table~\ref{tab:qnum}.
Similarly, $S-$wave modes, like the decays into $J/\psi \eta$ of the $1^{+-}$ states in Table~\ref{tab:qnum}, would 
be associated to a matrix element of the form $\langle J/\psi(\eta,p^\prime)\eta(q)|Y(p,\epsilon)\rangle={\cal F}$ where ${\cal F}$ must have 
dimensions of mass (one can set ${\cal F}=g M_Y$). Matrix elements of the form  $\langle J/\psi(\eta,p^\prime)\eta(q)|Y(p,\epsilon)\rangle={\cal N} (p\cdot \eta)(q\cdot \epsilon)$ would instead give the $D-$wave contribution. As for $P-$wave decays like those of the $0^{--}$ states we use the parameterization 
$\langle J/\psi(\epsilon,p^\prime) \eta(q)|0^{--}\rangle = {\cal H} (p\cdot \epsilon)$, where the dimensionless ${\cal H}$ is  ${\cal H}=A/\sqrt{2}$. The standard $\eta \eta^\prime$, $\omega\phi$ mixing schemes are used in the calculation.

As another possible decay mechanism consider the quark pair creation allowing a diquark-antidiquark system to decay into a tetraquark and a standard meson as in $[cs][\bar c\bar s]\to f_0(980)J/\psi$.
With quark  pair production one could also have baryon-antibaryon final states. Charmed baryons with strangeness ($\Xi_c$, ${\Omega^0}_c$) are anyway too heavy to be found in the decay products of  the hadrons in Table~\ref{tab:qnum}.  If we assume that the pair creation is regulated by $\alpha_s(m_c)$ rather than by $\alpha_s(\Lambda_{\rm QCD})$, we might infer that pair creation is less probable.
Along the same lines  we do not expect to have significant contributions form 6-fermion interactions, as those induced by instantons (see~\cite{thooft}). The reason is that the instanton Lagrangian is dumped by a factor 
$\exp(-8\pi/g^2)$ which turns out to be rather small at the mass scale of the charm quark where 
$\alpha_s(m_c)\sim 0.3$. In this sense the instanton interactions are mainly related to infrared physics.
Annihilation diagrams, also expected to be rather suppressed at the charm quark scale, could produce final states as $DD\pi$.

{\bf \emph{Experimental evidences}}.
Invariant mass spectra of several of the final states of interest have already been explored by experimental searches.

The most interesting match between predicted and observed states is in the 
$J/\psi \omega$ invariant mass spectrum, studied both by Belle~\cite{belle-omega} and 
by BaBar~\cite{babar-omega} in $B\to J/\psi \omega K$ decays. A state with mass $m_Y=3913\pm4$~MeV, 
according to the more accurate BaBar measurement, is observed to decay predominantly in this final state.
This paper shows how the $J^{PC}=0^{++}$ state, which decays predominantly into $J/\psi\omega$ and can
be produced in $B$ decays in pair with kaons if $L=0$, has a predicted mass of $3927$ MeV.

\begin{figure}[htb]
\begin{center}
\epsfig{height=3truecm, width=6truecm,figure=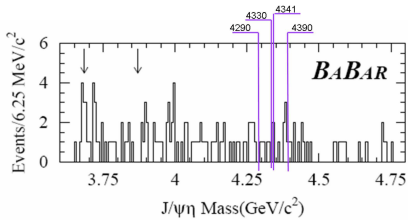}      
\caption{$J/\psi \eta$ events at BaBar.}
\label{ooo}
\end{center}
\end{figure}

The $J/\psi \eta$ invariant mass was studied by BaBar~\cite{babar-eta} in $B\to J/\psi \eta K$ decays,
and the resulting background subtracted distribution is reported in Fig.~\ref{ooo}.
Vertical lines refer to predicted mass value for the $J^{PC}=1^{--}$ states. Even if states with different 
$J^{PC}$ quantum number may decay in $J/\psi \eta$ final state, selection rules forbid $J^{PC}=0^{--}$ 
or $J^{PC}=1^{+-}$ states. 

The $\psi f_0(980)$ decay mode was instead studied when $f_0(980)$ decays into two pions, {\it i.e.} in the
$\psi\pi\pi$ final state, where $\psi$ can be either $J/\psi$ or $\psi(2S)$. Exotic mesons are searched 
in Initial State Radiation ({\it i.e.} $e^+e^-\to Y\gamma$ processes) and can therefore only be $J^{PC}=1^{--}$.
The published spectra~\cite{belle-psipipi,babar-psipipi} show several structures 
at $m=4260, 4350,$ and $4660$~MeV. Although only the latter one shows $\pi\pi$ invariant masses 
clearly consistent with an $f_0(980)$ production, it is interesting to notice that the invariant masses predicted
here for $J^{PC}=1^{--}$ states are in the same mass range.

Finally, possible exotic states decaying into $J/\psi\phi$ and $J/\psi\eta^{\prime}$ have been searched in 
Refs.~\cite{babar-phi} and~\cite{belle-etap} respectively, but no significant signal has been observed even integrating over the mass spectrum.

{\bf \emph{Conclusions}}. 
In this paper we have studied the consequences of allowing $[cs][\bar c\bar s]$ diquark-antidiquark
particles with different $J^{PC}$ quantum numbers. We present their spectrum and main decay modes in Table~\ref{tab:qnum}.
The comparison with existing data shows some hints of match between observed and predicted particles but significantly larger data 
are needed for conclusive statements. 

After our paper appeared, the CDF collaboration presented 
a $3.8~\sigma$ evidence of a narrow structure in $J/\psi \phi$ at about 4143~MeV~\cite{cdf}. 
As it is clear from Fig.~\ref{ooo}, we do not have anything close in our spectrum. 
Other interpretations of the CDF structure can be found in~\cite{news}.
In the same CDF paper, another peak in $J/\psi \phi$
is found with a significance $\lesssim 3~\sigma$   at about 4277~MeV. In this case we predict a $0^{-+}$ 
state at $4277$~MeV decaying into $J/\psi \phi$ in $P-$wave.

\bigskip 

\end{document}